\documentclass[aps, prb, letterpaper, 10pt, showpacs, twocolumn, superscriptaddress]{revtex4-2}
\usepackage{amssymb}
\usepackage{amsmath}
\usepackage{amsfonts}
\usepackage{graphicx}
\usepackage{bm}
\usepackage{xcolor}
\usepackage{natbib}
\usepackage{float}
\usepackage[colorlinks, linkcolor=blue, anchorcolor=blue, citecolor=blue]{hyperref}
\graphicspath{{figs/}{figsgaoerb/}}
\newcommand{\ket}[1]{\left\lvert #1 \right\rangle}
\newcommand{\fref}[2]{Fig.~\hyperref[#1]{\getrefnumber{#1}#2}}
\begin{document}
\title{Anti-crosstalk high-fidelity state discrimination for superconducting qubits}
\author{Zi-Feng Chen}
\affiliation{CAS Key Laboratory of Quantum Information, University of Science and Technology of China, Hefei, Anhui 230026, China}
\affiliation{CAS Center for Excellence and Synergetic Innovation Center in Quantum Information and Quantum Physics, University of Science and Technology of China, Hefei, Anhui 230026, China}
\author{Qi Zhou}
\affiliation{CAS Key Laboratory of Quantum Information, University of Science and Technology of China, Hefei, Anhui 230026, China}
\affiliation{CAS Center for Excellence and Synergetic Innovation Center in Quantum Information and Quantum Physics, University of Science and Technology of China, Hefei, Anhui 230026, China}
\author{Peng Duan}
\affiliation{CAS Key Laboratory of Quantum Information, University of Science and Technology of China, Hefei, Anhui 230026, China}
\affiliation{CAS Center for Excellence and Synergetic Innovation Center in Quantum Information and Quantum Physics, University of Science and Technology of China, Hefei, Anhui 230026, China}
\author{Wei-Cheng Kong}
\affiliation{Origin Quantum Computing Company Limited, Hefei, Anhui 230026, China}
\author{Hai-Feng Zhang}
\affiliation{CAS Key Laboratory of Quantum Information, University of Science and Technology of China, Hefei, Anhui 230026, China}
\affiliation{CAS Center for Excellence and Synergetic Innovation Center in Quantum Information and Quantum Physics, University of Science and Technology of China, Hefei, Anhui 230026, China}
\author{Guo-Ping Guo}
\email{gpguo@ustc.edu.cn}
\thanks{Corresponding author}
\affiliation{CAS Key Laboratory of Quantum Information, University of Science and Technology of China, Hefei, Anhui 230026, China}
\affiliation{CAS Center for Excellence and Synergetic Innovation Center in Quantum Information and Quantum Physics, University of Science and Technology of China, Hefei, Anhui 230026, China}
\affiliation{Origin Quantum Computing Company Limited, Hefei, Anhui 230026, China}

\date{\today}
\begin{abstract}
    Measurement for qubits plays a key role in quantum computation. Current methods for classifying states of single qubit in a superconducting multi-qubit system produce fidelities lower than expected due to the existence of crosstalk, especially in case of frequency crowding. Here, We make the digital signal processing~(DSP) system used in measurement into a shallow neural network and train it to be an optimal classifier to reduce the impact of crosstalk. The experiment result shows the crosstalk-induced readout error deceased by $100\%$ after a 3-second optimization applied on the 6-qubit superconducting quantum chip.
\end{abstract}
\maketitle
\section{Introduction}
	Single-shot qubit readout with high fidelity of a multi-qubit system is an essential feature in scalable quantum computation~\cite{article0}. For superconducting qubits, state-of-the-art readout is implemented by probing the qubit-state-dependent frequency shift of a readout resonator coupled to the qubit, which is based on dispersive interaction~\cite{article12, article3, article13}. Remarkable research progress has been made in rapid high-fidelity single-shot and multiplexed readout for circuit quantum electrodynamics~(CQED)~\cite{article11} systems of a few qubits~\cite{article16, article17, article1, article2, article18}. However, as systems scale up~\cite{article15,article27}, frequencies grow increasingly crowded, therefore each qubit could choose its frequency only from a limited range of frequency spectrum. Consequently, crosstalk induced by unwanted interactions between subsystems is aggravated by frequency crowding and it is critical to suppress the crosstalk to ensure optimal readout performance. From the perspective of data-processing, in frequency-multiplexed readout, the implementation of parallel mode-matched filters~\cite{article22} by weighted integration for each readout frequency effectively improves readout fidelity~\cite{article4, article6, article23}. But these methods focus on information extraction from individual frequency of each qubit with little consideration of part information leaked to other qubits. Thus, previous data processing flow is hard to give independent single-qubit information when crosstalk exists.

	Recently,machine learning~(ML) algorithms have been applied to many problems in quantum information such as measurement trajectories analysis~\cite{article8} and quantum dynamics reconstruction~\cite{article9}. But few of them are suitable for crosstalk mitigation in multiplexed readout scheme. In this work, we propose a simple but effective ML method based on DSP knowledge to build an anti-crosstalk classifier for state discrimination. In conventional DSP framework, the accuracy of state assignment for a single qubit would be perturbed by the state of the other qubits. Here, we use a trainable shallow neural network to mitigate the crosstalk.
	\section{Overview of the classifier architecture}
	In this section, we explain the typical DSP flow applied in qubits' readout and the crosstalk problem~\cite{article28} reflected in DSP domain. Then we introduce the method to build the architecture of our classifier based on the DSP chain.
	\begin{figure}[H]
		\centering
		\includegraphics[width=\columnwidth]{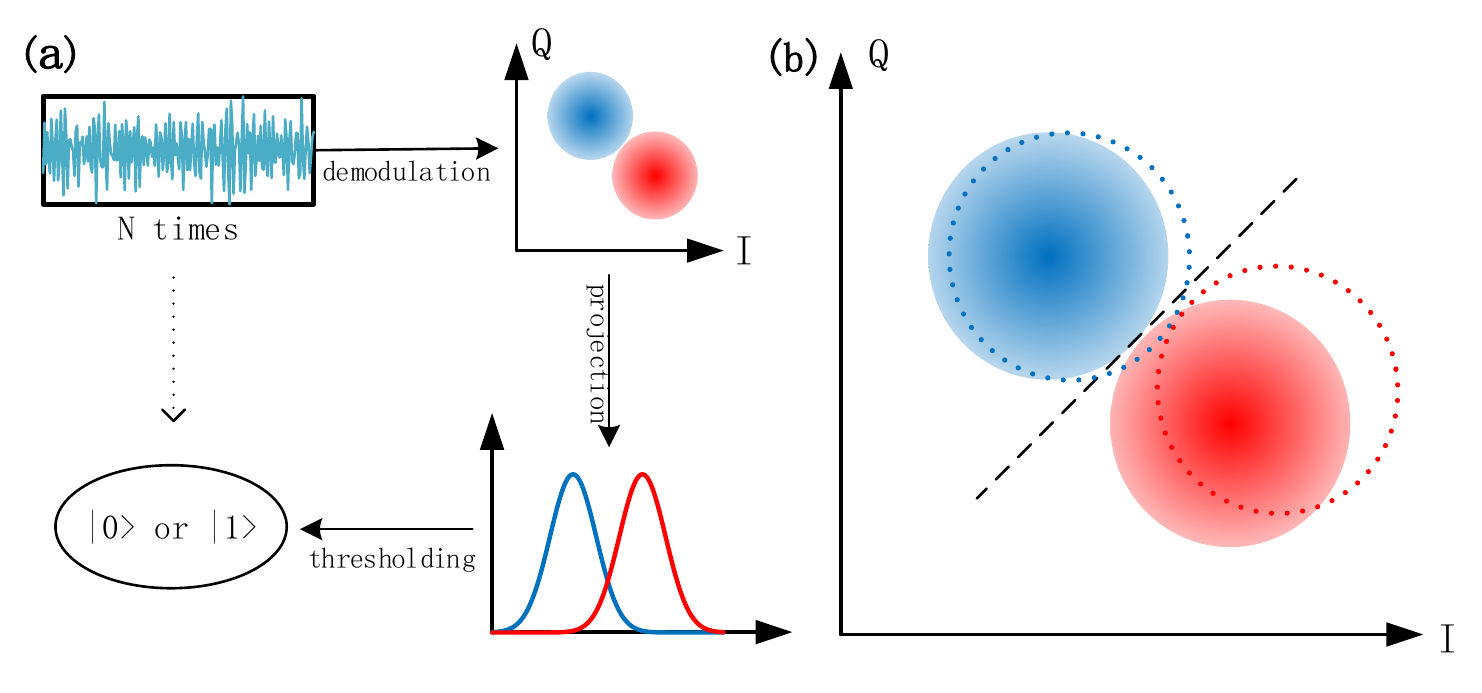}
		\caption{(a) Diagram of the DSP flow in readout. The raw signal is mapped onto I-Q plane by demodulation. Then projection and thresholding help to make a state decision.  (b) Shifts of I-Q clouds observed in a two-qubit system. The two solid circles are the I-Q clouds of Q1 with Q2 prepared in state $\ket{0}$ and the black dashed line is the optimum discriminant line found by support vector machine~(SVM) according to these two clouds. The dashed hollow circle stands for the shifts of the I-Q clouds when preparing Q2 in state $\ket{1}$ and the previous decision boundary is no longer optimum.}
		\label{fig1}
	\end{figure} 
	\subsection{Readout and crosstalk problem}
	In a typical DSP chain, a demodulation in a heterodyning scheme is performed after downconversion and digitalization of the reflected or transmitted tone from the readout resonator~\cite{article19}. This demodulation process involves multiple stages including mixing with an intermediate frequency~(IF), filtering and being integrated with a kernel which maybe an optimal demodulation envelope~\cite{article4}. This three-stage demodulation process would obtain in-phase~(I) and quadrature~(Q) components with encoded information about the qubit state. Repeating many times single-shot measurements for the target qubit would form two two-dimensional Gaussian statistical distributions on the I-Q plane. These two I-Q clouds would correspond to the two possible qubit states. Projecting the clouds onto the axis for which their relative separation in the complex plane is maximized produces a pair of one dimensional Gaussian curves. Further selection of an appropriate threshold acquires the optimized decision boundary for state discrimination. The DSP flow is depicted in \fref{fig1}{(a)}.

\begin{figure}[t]
        \centering
		\includegraphics[width=\columnwidth]{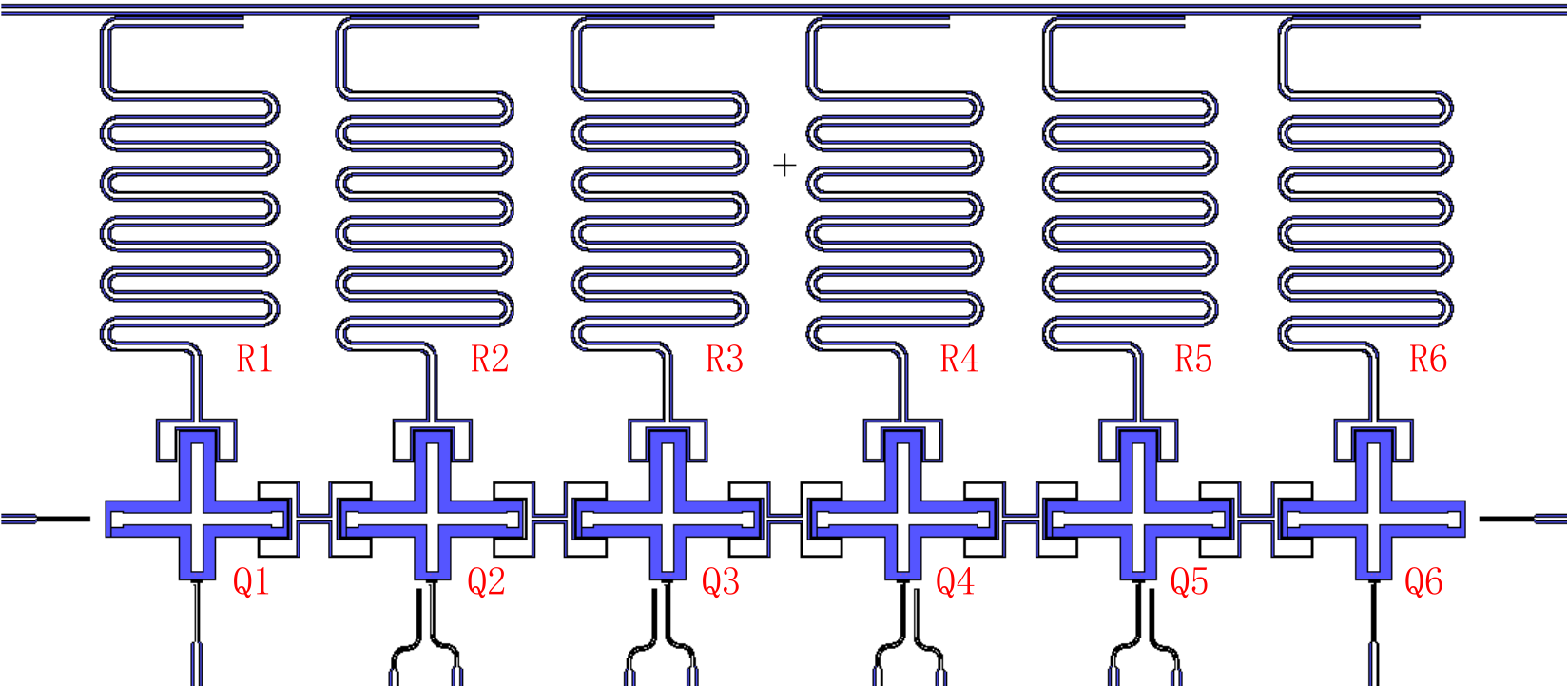}
		\caption{False-colored optical micrograph of our superconducting quantum chip. Each qubit is capacitively coupled to its neighbouring qubits and a readout resonator. The intermediate frequencies related to the readout resonators range from $\sim$500~MHz to $\sim$600~MHz. The frequency spacing between individual resonator frequencies is close to the designed value of 20~MHz.}
		\label{fig2}
	\end{figure}

	However, in practice, the result of demodulation may be influenced by not only one qubit. For example, in our quantum chip featuring six transmon qubits~\cite{article14,article7} as shown in Fig.~\ref{fig2}, readout resonator of Q1 could be affected by the state of Q2 to some extent. As a result, if we prepare Q2 in ground or excited state before measuring Q1, I-Q clouds would have an obvious shift as presented in \fref{fig1}{(b)}. We use $\ket{00}$, $\ket{01}$, $\ket{10}$ and $\ket{11}$ to denote the four possible basis states of $\ket{Q1,Q2}$. For Q1's readout, distinguishing $\ket{00}$ from $\ket{10}$ or $\ket{01}$ from $\ket{11}$ would both get near-optimal fidelity~($\boldsymbol F_1$ in Table~\ref{tab1}). But when attempting to distinguish $\{\ket{00},\ket{01}\}$ from $\{\ket{10},\ket{11}\}$ by a single classifier, fidelity has a significant decline~($\boldsymbol F_2$ in Table~\ref{tab1}). Here the fidelity is defined as the probability of correct assignment~\cite{article20} $F=[P(g|g)+P(e|e)]/2$, where $P(x|y)$ is the probability that the qubit prepared in state y is assigned as state x.
	\begin{table}[ht]
		\centering
		\begin{tabular*}{\linewidth}{c@{\extracolsep{\fill}}cc}
		\hline\hline
		Q2's state	            &0       &1        \\ \hline
		$\boldsymbol F_1$		&87.45\% &87.25\%  \\
		$\boldsymbol F_2$   	&86.25\% &86.80\%  \\
		\hline\hline
		\end{tabular*}
		\caption{Fidelities of Q1's readout.}
		\label{tab1}
	\end{table}
	\subsection{Architecture of our classifier}
\begin{figure}[h]
		\includegraphics[width=\columnwidth]{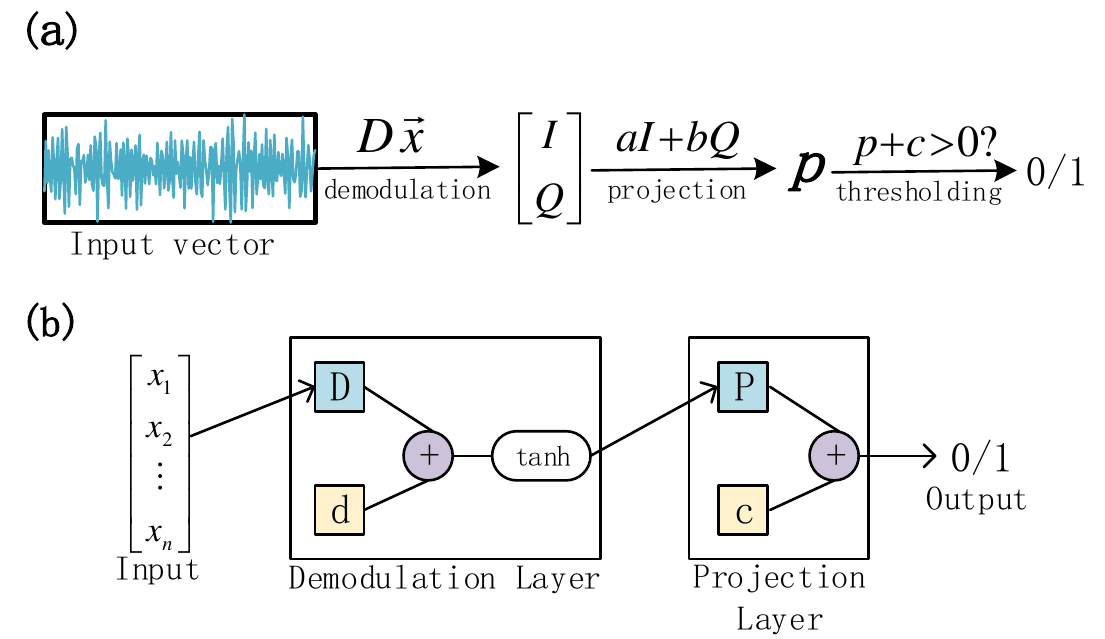}
		\caption{(a) Matrix form of the DSP chain of the single-qubit readout. (b) Network for state assignment based on the DSP system in (a).}
		\label{fig3}
	\end{figure}

	Three-stage demodulation pipeline has been mathematically proved to be equivalent to a single kernel integration stage~\cite{article5}. In other words, if we regard the signal sequence as a N-dimension column vector $\vec{x}$, the demodulation process is to left multiply the signal by a 2-by-N matrix D, mapping the signal to the I-Q plane:
	\begin{equation}\label{eq1}
		\begin{bmatrix}I\\Q\end{bmatrix}=D\vec{x}. 
	\end{equation}
	Suppose the function of discriminant line is $ax+by+c=0$, projection could also be regarded as a matrix manipulation:
	\begin{equation}\label{eq2}
		p=\begin{bmatrix}a&b\end{bmatrix}\begin{bmatrix}I\\Q\end{bmatrix}. 
	\end{equation}
	Thus, we could regard the data processing of readout as a linear algebra trick depicted in \fref{fig3}{(a)}. Slight modification of this architecture makes it into a two-layer neural network. As shown in \fref{fig3}{(b)}, addition of the biases d in the demodulation layer could move the I-Q points in I-Q plane and the tanh function takes nonlinearity into account. We define this network as the Quantum State Classifier~(QSC). Training QSC by gradient descent could optimize the fidelity.

	We introduce the optimization process of QSC with experimental data from our device. Here we select Q3 as the target qubit and mainly take Q2 and Q4's influence into consideration. Q3's state set $\ket{*0*}$~($\ket{*1*}$) has four subsets $\ket{000}$~($\ket{010}$), $\ket{001}$~($\ket{011}$), $\ket{100}$~($\ket{110}$) and $\ket{101}$~($\ket{111}$). The optimal integration time for Q3 is 200~ns, which is a trade-off between overlap error and $T_1$ events~\cite{article1}. Each single-shot signal has 320 time-points when sampled at 1.6~GS/s by an analog-to-digital converter~(ADC). We prepare 8000 records of single-shot readout, each 1000 initially prepared in one of the eight possible states $\ket{Q2,Q3,Q4}$. We initialize QSC with these samples according to conventional DSP theory. We initialize the weights of the demodulation layer using a weighted kernel \cite{article5}, the biases to the value making the two I-Q clouds symmetry about the origin. Inputing a small number of training samples to the demodulation layer would acquire the rough location of the two I-Q clouds and we write the perpendicular bisector of the line connecting their centers as $ax+by+c=0$. So that we could initialize the weights of the projection layer to $[a\ b]$, and the bias to $c$. Then we could start the training processing. We use an adaptive learning rate method~\cite{article24} and feature scaling of the input vector effectively accelerates the processing.
	\section{Result}
	We take Q3 as an instance merely with Q2 and Q4's influence considered. We use $S_1$, $S_2$, $S_3$ and $S_4$ to denote four state subsets~($\{\ket{000},\ket{010}\}$, $\{\ket{001},\ket{011}\}$, $\{\ket{100},\ket{110}\}$ and $\{\ket{101},\ket{111}\}$) of Q3 respectively. Under the framework of conventional DSP, an individual binary classifier~(based on SVM) trained with $S_i$~(i=1,2,3 or 4) after demodulation gets near-optimal fidelity on $S_i$ but fails to get optimal fidelities on $S_{j\neq i}$, see $\boldsymbol F_i$ in Table~\ref{tab2}. Further, another classifier trained with $\{S_1,S_2,S_3,S_4\}$ is hard to give consideration to different subsets simultaneously. Therefore, it fails to reach the expected fidelity~(mean value of $\{F_{ii}\}$ for i=1 to 4, $F_{ik}$ denotes the k-th value of $\boldsymbol F_i$), see $\boldsymbol F_5$ in Table~\ref{tab2}.
	\begin{table}[h]
		\centering
		\footnotesize
		\begin{tabular*}{\linewidth}{c@{\extracolsep{\fill}}cccc}
		\hline\hline
		$\ket{Q2,Q4}$	            &$\ket{00}$ &$\ket{01}$ &$\ket{10}$ &$\ket{11}$ \\ \hline
		$\boldsymbol F_1$		    &$\boldsymbol{89.15\%}$ &86.90\%    &84.75\%    &88.05\%    \\
		$\boldsymbol F_2$		    &86.90\%    &$\boldsymbol{89.05\%}$ &80.00\%    &87.55\%    \\
		$\boldsymbol F_3$		    &87.45\%    &79.60\%    &$\boldsymbol{88.00\%}$ &84.75\%    \\
		$\boldsymbol F_4$		    &88.10\%    &86.90\%    &85.55\%    &$\boldsymbol{89.20\%}$    \\ \hline
		$\boldsymbol F_5$		    &88.80\%    &87.55\%    &85.10\%    &88.65\%    \\ \hline
		$\boldsymbol F_6$		    &89.75\%    &90.20\%    &90.90\%    &90.40\%    \\
		\hline\hline
		\end{tabular*}
		\caption{Assignment fidelities of Q3's state. $\boldsymbol F_1$ to $\boldsymbol F_4$ stand for the results of considering only about one of the four possible states $\ket{Q2,Q4}$. $\boldsymbol F_5$ implies that it's hard to realize crosstalk-free state assignment in DSP scheme. $\boldsymbol F_6$ indicates the breakthrough of QSC.}
		\label{tab2}
	\end{table}
	
	Crosstalk makes information of qubits partly mixed with each other. Here we demonstrate the promotion in extracting useful information of QSC after training. In our experiment, 500 times training is enough for convergence in 3 seconds. The fidelities of the trained QSC on another 8000 records of single-shot readout are shown as $\boldsymbol F_6$ in Table~\ref{tab2}. The mean values $\boldsymbol\mu$ and the standard deviations $\boldsymbol\sigma$ of each $\boldsymbol F_i$~(for i=1 to 6) are shown in Table~\ref{tab3}. Here we use $\mu_i$ and $\sigma_i$ to denote the values belonging to $\boldsymbol F_i$.
	\begin{table}[h]
		\centering
		\footnotesize
		\begin{tabular*}{\linewidth}{c@{\extracolsep{\fill}}cccccc}
		\hline\hline
		$Fidelities$	         &$\boldsymbol F_1$ &$\boldsymbol F_2$ &$\boldsymbol F_3$ &$\boldsymbol F_4$ &$\boldsymbol F_5$ &$\boldsymbol F_6$      \\ \hline
		$\boldsymbol\mu$         &87.21\%           &85.87\%           &84.95\%           &87.44\%           &87.53\%           &$\boldsymbol{90.31\%}$ \\
		$\boldsymbol\sigma$		 &1.88\%            &4.02\%            &3.84\%            &1.57\%            &1.71\%            &$\boldsymbol{0.48\%}$  \\
		\hline\hline
		\end{tabular*}
		\caption{Mean values and standard deviations of fidelities. Fidelities are lower than expected and unstable under the framework of DSP. The trained QSC gets a more promoted performance}
		\label{tab3}
	\end{table}
	
	The mean value of $\{F_{ii}\}$~(for i=1 to 4) is $\mu_0=88.85\%$ and the standard deviation of $\{F_{ii}\}$ is $\sigma_0=0.57\%$). We define that a classifier is “crosstalk-free” for Q3 if the classifier satisfies the following two conditions~(its fidelities on four state subsets denoted as $\boldsymbol F$):
	\begin{enumerate}
	\item mean value of $\boldsymbol F$ is not less than $\mu_0$;
	\item standard deviation of $\boldsymbol F$ is not more than $\sigma_0$
	\end{enumerate}
	
	Under the framework of conventional DSP, independent binary classifiers after demodulation are hard to be crosstalk-free~($\boldsymbol F_1$ to $\boldsymbol F_5$). But the trained QSC is crosstalk-free with $\mu_6=90.31\%$ and $\sigma_6=0.48\%$. Moreover, $\mu_6>\mu_0$ means that other types of readout error have been simultaneously suppressed apart from crosstalk-induced error.

	In our experiment, QSCs have been also applied to the other five qubits and proved to be crosstalk-free. The performance of six QSCs for our quantum chip is shown in Table~\ref{tab4}. Here we define the fidelity loss in conventional DSP scheme as $f_{p_1}=\mu_5-\mu_0$ and the extra fidelity promotion of the trained QSC as $f_{p_2}=\mu_6-\mu_0$. For our device, QSC considering all the $2^6$ subsets of targeted qubit doesn't have obvious advantage over that only neighbouring qubits considered. Here we merely take neighbouring qubits into account for each qubit.
	\begin{table}[h]
		\centering
		\footnotesize
		\begin{tabular*}{\linewidth}{c@{\extracolsep{\fill}}cccccc}
		\hline\hline
		$Qubit$	         &$Q1$      &$Q2$      &$Q3$      &$Q4$      &$Q5$      &$Q6$      \\ \hline
		$\sigma_0$       &1.08\%    &1.89\%    &0.57\%    &1.36\%    &1.72\%    &0.36\%    \\ \hline
		$\sigma_5$       &1.37\%    &1.63\%    &1.71\%    &0.93\%    &2.26\%    &0.78\%    \\
		$f_{p_1}$		 &-0.92\%   &-0.36\%   &-1.32\%   &-0.44\%   &-3.06\%   &-0.44\%    \\ \hline
		$\sigma_6$       &0.95\%    &1.52\%    &0.48\%    &0.76\%    &1.51\%    &0.19\%    \\
		$f_{p_2}$		 &0.62\%    &1.35\%    &1.46\%    &0.88\%    &0.63\%    &0.22\%    \\
		\hline\hline
		\end{tabular*}
		\caption{Performance of trained QSCs. All of them are crosstalk-free and get a positive extra promotion of readout fidelities with higher stability than those in DSP scheme.}
		\label{tab4}
	\end{table}
	\section{Discussion and outlook}
\begin{figure}[t]
		\includegraphics[width=\linewidth]{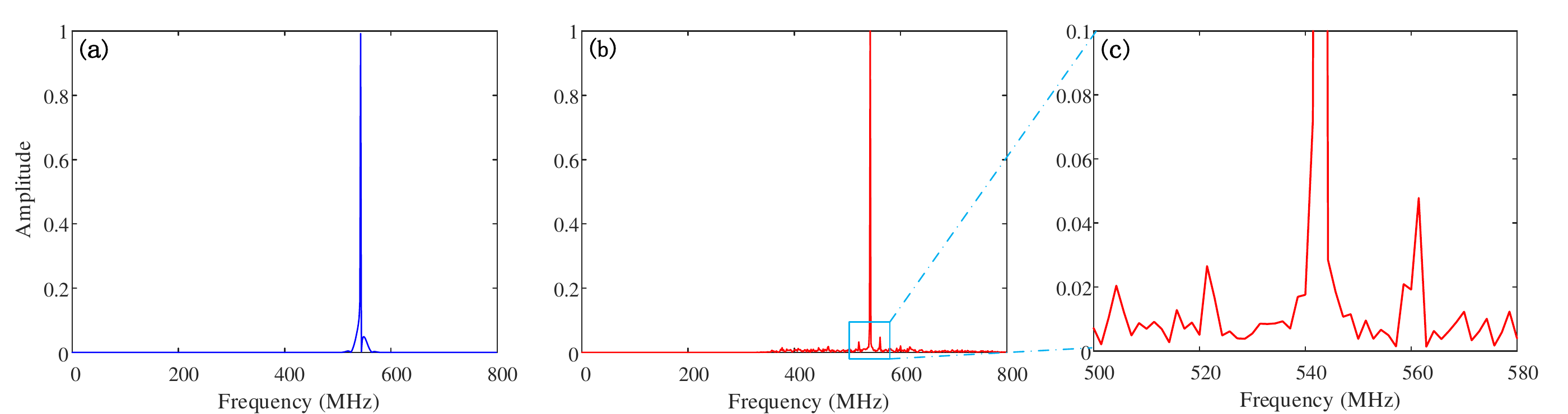}
		\caption{Amplitude spectrum of the first row of D in the first Nyquist zone. Similar change happened with the second row relevant to the Q quadrature is not shown. (a) Before training. The main frequency component is merely related to the target qubit Q3. (b) After training. There are two small peaks relevant to Q2 and Q4. (c) Enlarged view of the three peaks in the spectrum.}
		\label{fig4}
	\end{figure}

\begin{figure}[h]
		\includegraphics[width=\columnwidth]{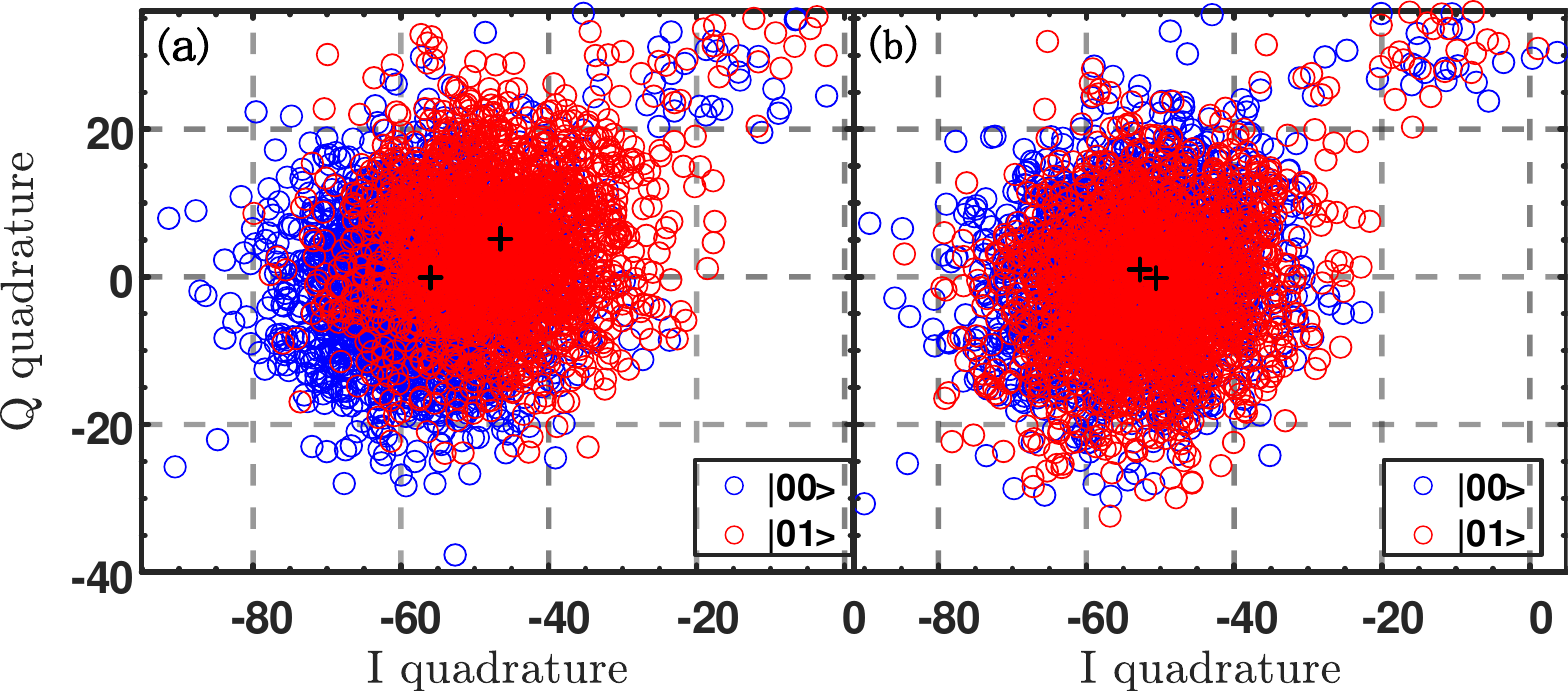}
		\caption{I-Q clouds correspond to ground state of Q3 involving Q4's influence. $\ket{Q3,Q4}$ stands for the state. Black crosses mark the centers of clouds. (a) Demodulated by conventional weighted kernel.  (b) Demodulated by trained weighted kernel.}
		\label{fig5}
	\end{figure}

	In conclusion, we have demonstrated performance of QSC in suppression of readout crosstalk. For instance, the trained QSC for Q3 gets consistent optimal fidelities on any subsets of Q3's state. We infer that QSC acquires the ability to regather the leaked information to other frequency components by learning. We could evaluate this speculation by observing the amplitude spectrum's change of the weighted integration kernel of the demodulation, see Fig.~\ref{fig4}. In the amplitude spectrum, the conventional weighted kernel has a single peak located at the IF of Q3's readout resonator. But after training, two small peaks appear at the IFs of the two neighbouring resonators. This implies, to some extent, that the trained QSC gets access to purify the mixed information. The promoted performance of the demodulation layer of QSC is shown in Fig.~\ref{fig5}. The shift of I-Q clouds caused by crosstalk has been almost eliminated after training. In this work, QSC acquires additional optimization of readout fidelity. We look forward to further improvement of QSC architecture and its better performance in quantum state discrimination.

	In frequency-multiplexed readout scheme, QSC could be a unified optimal classifier with effective suppression of information from untargeted qubits. For our six-qubit device, reaching the optimal fidelity of single-shot readout needs merely six independent QSCs when crosstalk exists. Compared with other sophisticated ML methods applied in quantum information~\cite{article25, article26, article8, article9}, the much simpler architecture with DSP-based initialization has a great advantage over training. In addition, QSC is scalable for system with larger numbers of qubits. And it could be transplanted to field-programmable gate array~(FPGA) platform for realtime feedback control~\cite{article21, article5, article10}.

	Moreover, it implies that crosstalk problems could be solved in DSP domain to a degree. For hardware design, it is in favour of selecting smaller frequency spacing under limitation of the bandwidth of the amplifiers so that more qubits are allowed in a single measurement channel.
	\section{Acknowledgments}
	We thank \emph{OriginQ,Inc.} for hardware support. The quantum chip we employed is \href{http://originqc.com.cn/en/website/productDetail.html?id=206&bannerId=116}{KF~C6-130}. This work was supported by the National Key Research and Development Program of China (Grant No.~2016YFA0301700), the National Natural Science Foundation of China (Grants No.~12034018,11625419), the Strategic Priority Research Program of the CAS (Grant No.~XDB24030601), the Anhui initiative in Quantum Information Technologies (Grants No.~AHY080000). This work was partially carried out at the USTC Center for Micro and Nanoscale Research and Fabrication.

\end{document}